\documentclass[11pt,a4paper]{article}
\pdfoutput=1
\usepackage{jheppub}

\usepackage{float}

\usepackage[normalem]{ulem}

\title{
Enhancing the Large Hadron Collider Sensitivity to Charged and Neutral Broad Resonances of New Gauge Sectors
}

\author[a]{J. Fiaschi,}
\author[b,c,d]{F. Giuli,}
\author[b,e,f]{F. Hautmann}
\author[g,h]{and S. Moretti}

\affiliation[a]{Department of Mathematical Sciences, University of Liverpool, Liverpool L69 3BX, United Kingdom}
\affiliation[b]{CERN - European Organization for Nuclear Research,
Espl. des Particules 1, 1211 Meyrin, Switzerland}
\affiliation[c]{University of Rome Tor Vergata and INFN, Sezione di Roma 2, Via della Ricerca Scientifica 1, 00133 Roma}
\affiliation[d]{Brandeis University, 415 South St, Waltham, MA 02453, United States}
\affiliation[e]{Elementaire Deeltjes Fysica, Universiteit Antwerpen, B 2020 Antwerpen}
\affiliation[f]{Theoretical Physics Department, University of Oxford, Oxford OX1 3PU}
\affiliation[g]{School of Physics and Astronomy, University of Southampton, Highfield, Southampton SO17 1BJ, United Kingdom}
\affiliation[h]{Department of Physics and Astronomy, Uppsala University, Box 516, SE-751 20 Uppsala, Sweden}

\emailAdd{fiaschi@liverpool.ac.uk}
\emailAdd{francesco.giuli@cern.ch}
\emailAdd{hautmann@thphys.ox.ac.uk}
\emailAdd{stefano.moretti@cern.ch}

\abstract{
In scenarios beyond the Standard Model (BSM) characterised by charged ($W^\prime$) or neutral ($Z^\prime$) massive gauge bosons with large width, resonant mass searches are not very effective, so that one has to exploit the tails of the mass distributions measured at the Large Hadron Collider (LHC). In this case, the LHC sensitivity to new physics signals is influenced significantly by systematic uncertainties associated with the Parton Distribution Functions (PDF), particularly in the valence quark sector relevant for the multi-TeV mass region. 
As a BSM framework featuring such conditions, we consider the 4-Dimensional Composite Higgs Model (4DCHM), in which multiple $W^\prime$ and $Z^\prime$ broad resonances are present, with strongly correlated properties. By using the QCD tool \texttt{xFitter}, we study the implications 
on $W^\prime$ and $Z^\prime$ searches in Drell-Yan (DY) lepton decay channels that follow from the reduction of PDF uncertainties obtained through combining high-statistics precision measurements of DY lepton-charge and forward-backward asymmetries.
We find that the sensitivity to the BSM states is greatly increased with respect to 
the case of base PDF sets, thereby enabling one to set more stringent limits on (or indeed discover) such new particles, both independently and in correlated searches.
}

\preprint{LTH 1277}


\begin{document} 
\maketitle
\flushbottom

\section{Introduction}

Searches for heavy spin-1 $W^\prime$~\cite{Mohapatra:1974gc} and $Z^\prime$~\cite{Langacker:2008yv} bosons, corresponding to extra gauge symmetries Beyond the Standard Model (BSM), constitute one of the classic avenues to explore potential signatures of new physics at the Large Hadron Collider (LHC)~\cite{CidVidal:2018eel} as well as at future colliders~\cite{Golling:2016gvc}. 
In the case of BSM scenarios with narrow vector resonances, one can rely on traditional ``bump search" analyses based on the Breit-Wigner (BW) line shape --- see, e.g., the recent analyses~\cite{ATLAS:2019erb,ATLAS:2019lsy,CMS:2018ipm} in leptonic decay channels (i.e., in so-called Drell-Yan (DY) processes).
In the case of BSM scenarios featuring vector resonances with large widths, in contrast, alternative experimental strategies are needed --- see, e.g., the discussion in~\cite{CMS:2021ctt} --- which are much more dependent than bump searches on the modelling of the dominant production processes~\cite{Accomando:2019ahs}, for both signal and background. 

In this case, non-perturbative Quantum Chromodynamics (QCD) effects embodied by the 
Parton Distribution Functions (PDFs) in the initial state of hadronic collisions represent one of the main sources of theoretical systematic uncertainties, which affects the experimental sensitivity to test BSM scenarios, thus influencing the potential of experimental searches for discovering or setting exclusion bounds on heavy BSM vector bosons.
In particular, in the multi-TeV region of $W^\prime$ and $Z^\prime$ masses defined by the current LHC exclusion limits~\cite{ATLAS:2019erb,ATLAS:2019lsy,CMS:2018ipm,CMS:2021ctt}, quark distributions in the valence sector, for moderate to large values of longitudinal momentum fractions $x$, dominate the PDF systematics. 

In Ref.~\cite{Fiaschi:2021okg}, we have proposed new approaches to improve the valence quark systematics, based on combining high-precision measurements of Charged Current (CC) and Neutral Current (NC) DY asymmetries in the mass region near the Standard Model (SM) vector boson poles. We have exploited the sensitivity of the NC forward-backward asymmetry $A_{\rm{FB}}$ to the charge-weighted linear combination $ (2/3) u_V + (1/3) d_V$ of up-quark and down-quark valence distributions~\cite{Accomando:2019vqt} and the sensitivity of the CC lepton-charge asymmetry $A_W$ to the difference $ u_V - d_V$.
By a quantitative ``profiling'' analysis using the \texttt{xFitter} platform~\cite{Alekhin:2014irh}, we have illustrated the complementary constraints provided by the two asymmetries on linearly independent combinations of $u$ and $d$ quark distributions. The constraints from the $A_{\rm{FB}}$ and $A_W$ combination, examined in the two projected luminosity scenarios of 300 fb$^{-1}$ (for the LHC Run 3) and of 3000 fb$^{-1}$ (for the High-Luminosity LHC (HL-LHC)~\cite{Gianotti:2002xx,CidVidal:2018eel}), turn out to improve the relative PDF uncertainties by up to around 20\% \cite{Fiaschi:2021okg} in the region of the invariant and transverse mass spectra between 2 TeV and 6 TeV, in which evidence for $W^\prime$ and/or $ Z^\prime$ states with large widths could first be observed. 

Motivated by this result, in this work we analyze the impact of improving the PDF systematics with the method of Ref.~\cite{Fiaschi:2021okg} on the 
experimental sensitivity of forthcoming $W^\prime$ and $ Z^\prime$ searches at the LHC, focusing on scenarios characterised by multiple $W^\prime $ and $ Z^\prime$ broad resonances and by interference effects of the heavy bosons 
with each other and with SM gauge bosons in the CC and NC channels, respectively. To this end, we consider new gauge sectors in strongly-coupled models of electroweak symmetry breaking~\cite{Panico:2015jxa} based on composite Nambu-Goldstone Higgs~\cite{Kaplan:1983fs,Kaplan:1983sm}. 

More precisely, we take the 4-Dimensional Composite Higgs Model (4DCHM) realization~\cite{DeCurtis:2011yx} of the minimal composite Higgs model of Ref.~\cite{Agashe:2004rs}.
We characterize the parameter space of the model using two parameters, the compositeness scale $f$ and the coupling $g_\rho$ 
of the new resonances. In terms of these parameters, the resonance mass scale is of order $M \sim f g_\rho$~\cite{Panico:2015jxa,Giudice:2007fh}. 
The 4DCHM scenario contains five neutral $Z^\prime$ bosons, $Z_i$ with $ i = 1 , \dots, 5$, and 
three charged $W^\prime$ bosons, $W_j$ with $ j = 1 , \dots, 3$. Of these, 
the $Z_1$, $Z_4$ and $W_1$ states do not couple to leptons and light quarks, so that 
they do not contribute to DY production and decay, while the $Z_5$ and $W_3$ states are too heavy to give non-negligible contributions.
Therefore, our analysis deals with the $Z_2, Z_3$ and $W_2$ bosons of the 4DCHM, as in the implementation~\cite{Barducci:2012kk} of the model.
In the 4DCHM the masses are correlated across the NC and CC sectors. We thus explore the LHC sensitivity to these BSM states using the improved PDFs~\cite{Fiaschi:2021okg} and 
studying the potential of correlated analyses across the NC and CC channels.

Our study will exploit the multi-resonant profile of the mass spectrum, and in particular the following important feature.
The strong interference effects between the BSM resonances themselves and between the BSM and SM states are responsible for the presence of a statistically significant depletion of events that appears below the BW (Jacobian) peaks in the invariant (transverse) mass distribution in the NC (CC) process~\cite{Accomando:2016mvz, Accomando:2019ahs, Fiaschi:2021okg}.
This effect is particularly relevant in the NC case, where the cumulative interference effects due to the $Z_2$ and $Z_3$ resonances lead to the appearance of a pronounced dip. A similar, though less marked, behavior is present in the CC case as well.
As already advocated in~\cite{Accomando:2016mvz, Accomando:2019ahs, Fiaschi:2021okg}, it is possible to define the significance of the depletion of events in a manner similar to that for the excess of events of the peaks, which can be used to extract model dependent exclusion and discovery limits in the model's parameter space.
In the following, we will present such limits resulting from the analysis of the spectra for either the peak or the dip. We will show that the latter receives an especially considerable benefit from the improvement in the PDF determination, since it occurs in invariant (transverse) mass regions in which 
the systematic PDF uncertainty is comparable to the statistical precision of the measurements.

The paper is organized as follows. In Sec.~\ref{sec:framework} we give a brief description of the computational framework, consisting of the implementation of the 4DCHM and of the improved PDFs following from the $A_{\rm{FB}}$ and $A_W$ asymmetry profiling with \texttt{xFitter}. In Secs.~\ref{sec:NC} and \ref{sec:CC} we present the results of our study within the 4DCHM for the neutral and charged sector, respectively, and discuss the impact of the improved PDFs on searches for additional gauge bosons in the two sectors, 
including correlated analyses across NC and CC. We present our conclusions in Sec.~\ref{sec:conc}.

\section{Computational framework}
\label{sec:framework}

\subsection{Implementation of the 4DCHM}

We describe the parameter space of the 4DCHM in terms of two parameters, the compositeness scale $f$ and the gauge coupling $g_\rho$ of the extra symmetry group~\cite{Panico:2015jxa,DeCurtis:2011yx}.
In the first approximation, the masses of the lightest resonances $M_{Z_2}$ and $M_{Z_3} \simeq M_{W_2}$ are of the order of $f g_\rho / \cos\psi$ and $f g_\rho / \cos\theta$ respectively, where $\psi \sim g_Y / g_\rho$, $\theta \sim g / g_\rho$ and $g$ and $g_Y$ are the gauge couplings to the SM gauge groups $SU(2)_L$ and $U(1)_Y$ respectively. Furthermore, additional corrections to the BSM gauge boson masses of order $v^2 / f^2$, with $v$ the SM Higgs vacuum expectation value, arise upon electroweak symmetry breaking. 

The numerical implementation of the 4DCHM in our framework follows the description given in~\cite{Barducci:2012kk}, already used in Ref.~\cite{Accomando:2016mvz} for the phenomenological analysis of its neutral gauge sector.
The same dedicated \texttt{Mathematica} program described in Ref.~\cite{Barducci:2012kk} has been used to obtain the chiral couplings to fermions of the heavy gauge bosons, which in turn have been included in a dedicated \texttt{C++} code for the computation of the invariant and transverse mass distributions used for the analysis in the NC and CC channels, respectively.
The \texttt{Mathematica} code uses the $G_\mu$ scheme~\cite{Hollik:1988ii} for the EW input parameters. The explicit values for the relevant parameters in our analysis are: $\alpha_{em}$ = 1/128.9, $M_Z$ = 91.1876 GeV, $\Gamma_Z$ = 2.441 GeV, $M_W$ = 80.370 GeV, $\Gamma_W$ = 2.085 GeV and the Fermi constant $G_F$ = 1.16639 $\cdot$ 10$^{-5}$ GeV$^{-2}$; the value of $\sin^2\theta_W$ is derived from the tree-level relation and no running of $\alpha_{em}$ is considered.
The allowed parameter space points obtained varying the model parameters $f$ and $g_\rho$ are required to predict the top quark mass in the interval 165 GeV $\leq m_t \leq$ 175 GeV, the bottom quark mass in the interval 2 GeV $\leq m_b \leq$ 6 GeV and the Higgs boson mass in the interval 124 GeV $\leq m_H \leq$ 126 GeV. The different levels of accuracy required for the three masses are due to the fact that the former two are computed at lowest order while the latter one is calculated through higher orders, as described in \cite{Barducci:2012kk}.

For the analysis of the neutral gauge sector, the scan over the parameter plane ($f$, $g_\rho$) has been obtained varying $f$ between 0.5 TeV and 5.5 TeV in steps of 0.1 TeV, while $g_\rho$ was varied between 1 and 6 in steps of 0.1.
The invariant mass distributions of the di-lepton spectra have been generated taking 1200 points equally separated with 40 GeV $\leq M_{\ell\ell} \leq$ 13 TeV for each realization of the 4DCHM neutral resonances in the parameter space.
The scan for the analysis of the charged gauge sector has been generated varying $f$ between 0.5 TeV and 5.5 TeV in steps of 0.2 TeV, while $g_\rho$ was varied between 1 and 6 in steps of 0.2.
The transverse mass distributions of the lepton-neutrino spectra have been produced computing 109 points equally separated with 100 GeV $\leq M_T \leq$ 11 TeV for each realization of the 4DCHM charged resonances in the model parameter space.

Finally, we note that, as discussed in \cite{Barducci:2012kk}, the additional parameters in the strong sector (the new fermion masses and mixings) can be fine-tuned in order to produce the required values of the various $Z^\prime$ and $W^\prime$ widths, which we fix to 20\% of their masses in all cases.
Sensitivity studies in composite Higgs models comparing dilepton with heavy-quark channels in broad resonance searches may be found e.g. in~\cite{Liu:2019bua, Jung:2019iii}.

\subsection{PDF Profiling}
The results of this paper are obtained using the ``improved PDFs" constructed according to the approach illustrated in Ref.~\cite{Fiaschi:2021okg}. This is based on the following: i) exploiting DY production for NC and CC channels in the mass region near the SM vector boson peak, characterised by high statistics; ii) using the information encoded in the DY asymmetries of the NC~\cite{Accomando:2019vqt,Abdolmaleki:2019ubu,Accomando:2018nig,Accomando:2017scx,Yang:2021cpd,Fu:2020mxl} and CC channels, associated with the SM vector boson left-handed and right-handed polarizations, to reduce quark PDF uncertainties.\footnote{The vector boson longitudinal polarization can be used to constrain gluon distributions in gluon fusion processes, see~\cite{Amoroso:2020fjw}.} 

In this approach the main point is thus to use experimental information on the DY vector boson polarization in the region near the peak, on top of the information on the unpolarized production cross section, to obtain PDFs with reduced uncertainties compared to the standard, or base, PDFs.
Out of the full structure of DY angular distributions~\cite{Abdolmaleki:2019ubu} associated with vector-boson polarized contributions, Ref.~\cite{Fiaschi:2021okg} picks out the NC and CC asymmetry distributions, as these provide linearly independent combinations of $u$ and $d$ valence quark distributions, which dominate the uncertainties for moderate to large values of longitudinal momentum fractions $x$. These are relevant especially for production processes in the high-mass region. 

To carry out this analysis, Ref.~\cite{Fiaschi:2021okg} employs the profiling technique~\cite{Paukkunen:2014zia} within the open-source \texttt{xFitter} platform~\cite{Alekhin:2014irh}. 
The use of profiling in the approach~\cite{Fiaschi:2021okg} differs from the in-situ profiling and reweighting analyses (e.g.~\cite{CMS:2018ktx,ATLAS:2018gqq}), as the purpose here is to look for new measurements capable of providing high sensitivity to PDFs with low theoretical and experimental systematics while controlling correlations.

\begin{figure}
\begin{center}
\includegraphics[width=0.45\textwidth]{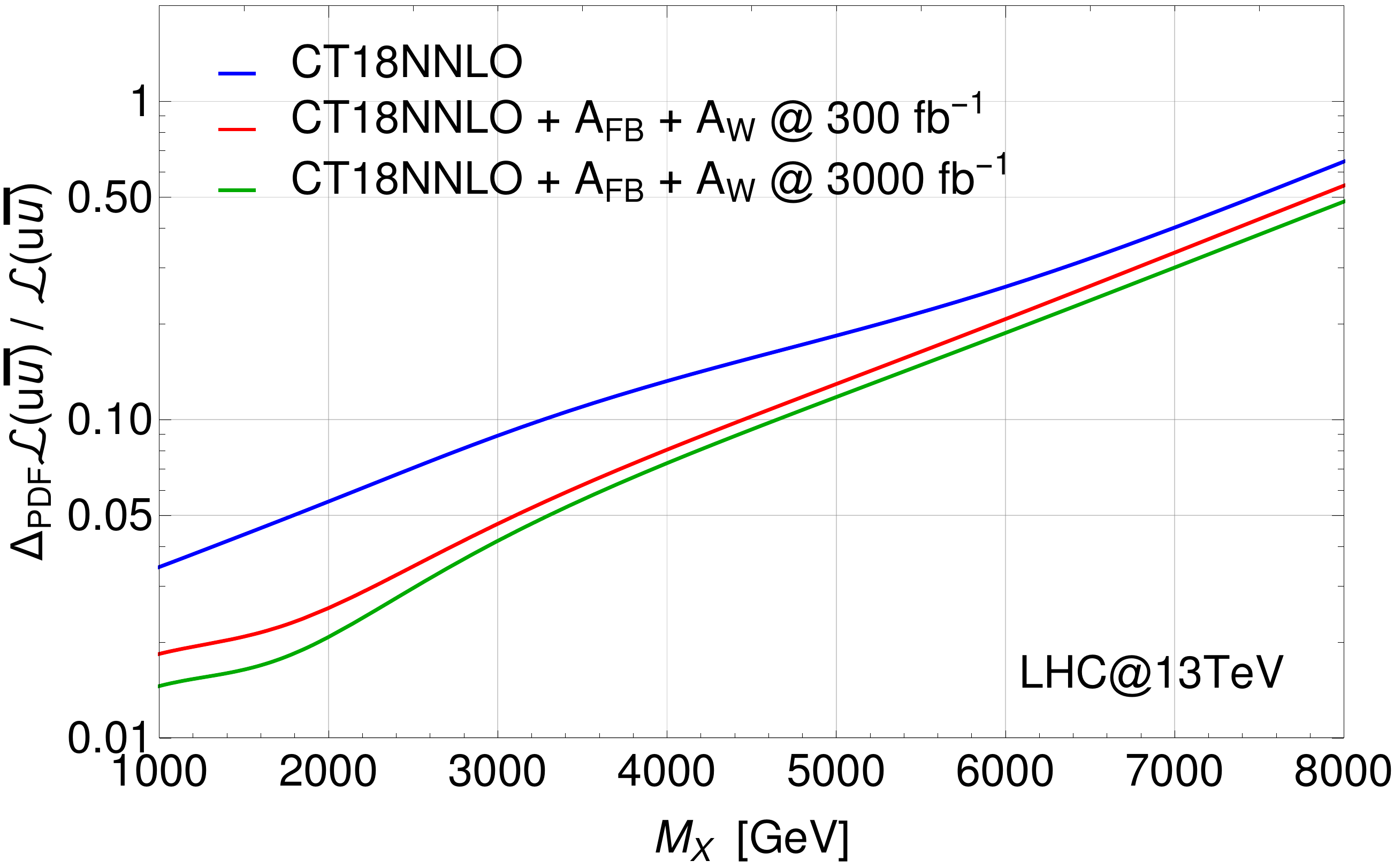}
\includegraphics[width=0.45\textwidth]{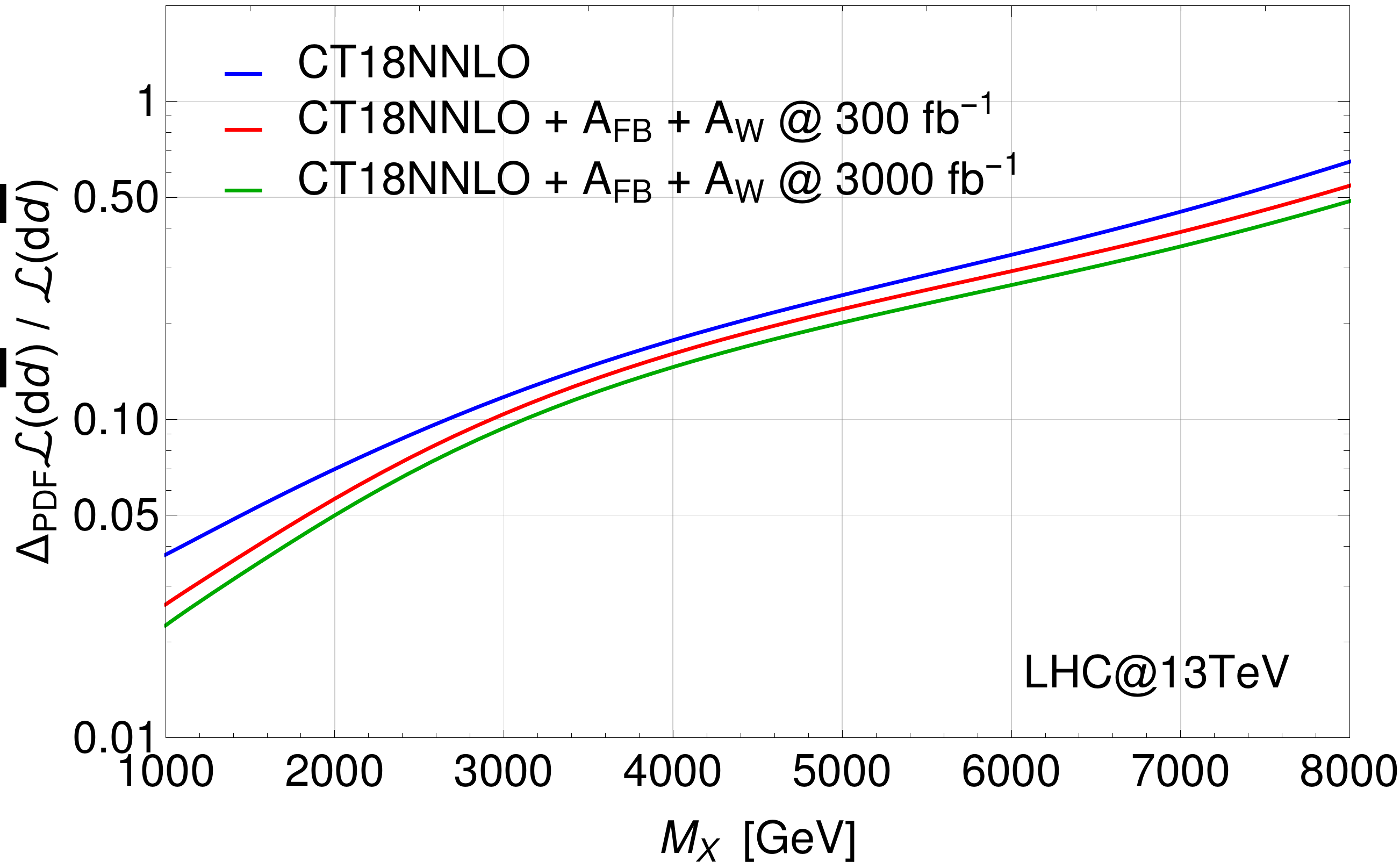}
\end{center}
\caption{Relative PDF error on the $u\bar{u}$ (left) and $d\bar{d}$ (right) parton luminosities, for the baseline CT18NNLO PDF set (blue) and the ``improved'' sets profiled using $A_{FB}$ and $A_W$ pseudodata with statistical uncertainty corresponding to 300 fb$^{-1}$ (red) and 3000 fb$^{-1}$ (green)}
\label{fig:Parton_Lumi}
\end{figure}

The NC observable $A_{\rm{FB}}$ has been implemented as a function of the invariant mass of the dilepton pair in the final state ($m_{\ell\ell}$) following the work done in Ref.~\cite{Accomando:2019vqt}.
The implementation of the CC observable $A_{W}$ as a function of the pseudorapidity of the charged lepton ($\eta_{\ell}$) has been similarly conducted, with the inclusion of the fiducial cuts reported in Ref.~\cite{ATLAS:2019fgb}, using NLO QCD predictions computed with \texttt{MadGraph5aMC@NLO}~\cite{Alwall:2014hca} as well as NNLO $k$-factors obtained from \texttt{DYNNLO}~\cite{Catani:2009sm}.
NLO EW corrections are not included, based on the studies carried out in~\cite{Accomando:2019vqt} 
which indicated that their impact on the profiling at the weak boson mass scale is not large.
With this method, Ref.~\cite{Fiaschi:2021okg} presents two sets of improved PDFs, obtained from profiling of the CT18 NNLO PDF set~\cite{Hou:2019efy}, for two different integrated luminosity scenarios, namely 300~fb$^{-1}$ (LHC Run 3) and 3000~fb$^{-1}$ (HL-LHC~\cite{CidVidal:2018eel}).
In Fig.~\ref{fig:Parton_Lumi} we show the improvement on the PDF determination after the profiling in the two scenarios by plotting the relative PDF uncertainties on the $q\bar{q}$ parton luminosities containing the contribution from $u$-like and $d$-like quarks.

Recent analyses have pointed out that including data containing contamination from BSM contributions in PDF fits assuming purely SM predictions would lead to biased estimations on new physics constraints~\cite{Greljo:2021kvv,CMS:2021yzl}.
In our analysis, the pseudodata employed in the profiling contains only SM contribution and it is centered around the SM vector boson peaks.
Since we consider TeV scale BSM resonances, the new physics contribution to the $A_{FB}$ at the EW scale, which is due to the interference between SM and BSM process, is small~\cite{Accomando:2015cfa} and a similar conclusion holds for the CC differential cross section distribution~\cite{Accomando:2011eu}.

In the next sections, we will study how the use of these improved PDFs can enhance the experimental sensitivity to broad $W^\prime$ and $ Z^\prime$ resonances in the 4DCHM. The data used to profile the PDF is chosen consistently with respect to the LHC luminosity stage that is considered in each scenario.
Finally, we select two 4DCHM benchmarks, denoted by A and B, for further analysis.
The details of the parameters of the benchmarks together with the respective heavy gauge bosons masses are listed in Tab.~\ref{tab:AB_bench}.

\section{Limits on the model parameter space from Neutral DY}
\label{sec:NC}

The neutral gauge sector of the 4DCHM model features three extra non-inert heavy resonances~\cite{Barducci:2012kk,Accomando:2016mvz}.
The first two resonances $Z_2$ and $Z_3$ are generally close in mass, with a separation of $\mathcal{O}$(100 GeV).
The other resonance $Z_5$ is generally much less coupled to SM matter and $\mathcal{O}$(1 TeV) heavier.
Its effects on the invariant mass distribution are negligible and we will not consider their phenomenology in the following.

We fix the width of the neutral resonances to $\Gamma_{Z^\prime} / M_{Z^\prime} =$ 20\%.
The overlapping BW functions of the $Z_2$ and $Z_3$ resonances, together with their strong interference effects (mutual and with the SM background), generate non-standard profiles for the invariant mass distributions, which will appear more like a shoulder over a smooth background.
We will therefore adopt a ``counting strategy" analysis, choosing optimal integration limits to maximize the sensitivity to the specific chosen BSM realization.

The overall significance is evaluated combining the sensitivities of the electron and muon channels, including their acceptances and efficiencies~\cite{CMS:2014lcz}, wherein the statistical and PDF uncertainties are summed in quadrature and other sources of systematic error are assumed to be negligible.

\begin{table}
\begin{center}

\begin{tabular}{|c||c|c|c|c|c|c|}
\hline
Benchmark & $f$ [TeV] & $g_\rho$ & $M_{Z_2}$ [TeV] & $M_{Z_3}$ [TeV] & $M_{W_2}$ [TeV] & $M_{W_3}$ [TeV]\\
\hline
A & 3.9 & 1.2 & 5.16 & 5.56 & 5.56 & 6.62\\
\hline
B & 1.5 & 2.2 & 3.39 & 3.45 & 3.45 & 4.67\\
\hline
\end{tabular}
\end{center}
\caption{Parameters for the benchmarks A and B and their heavy gauge bosons masses.}
\label{tab:AB_bench}
\end{table}

\begin{figure}
\begin{center}
\includegraphics[width=0.4\textwidth]{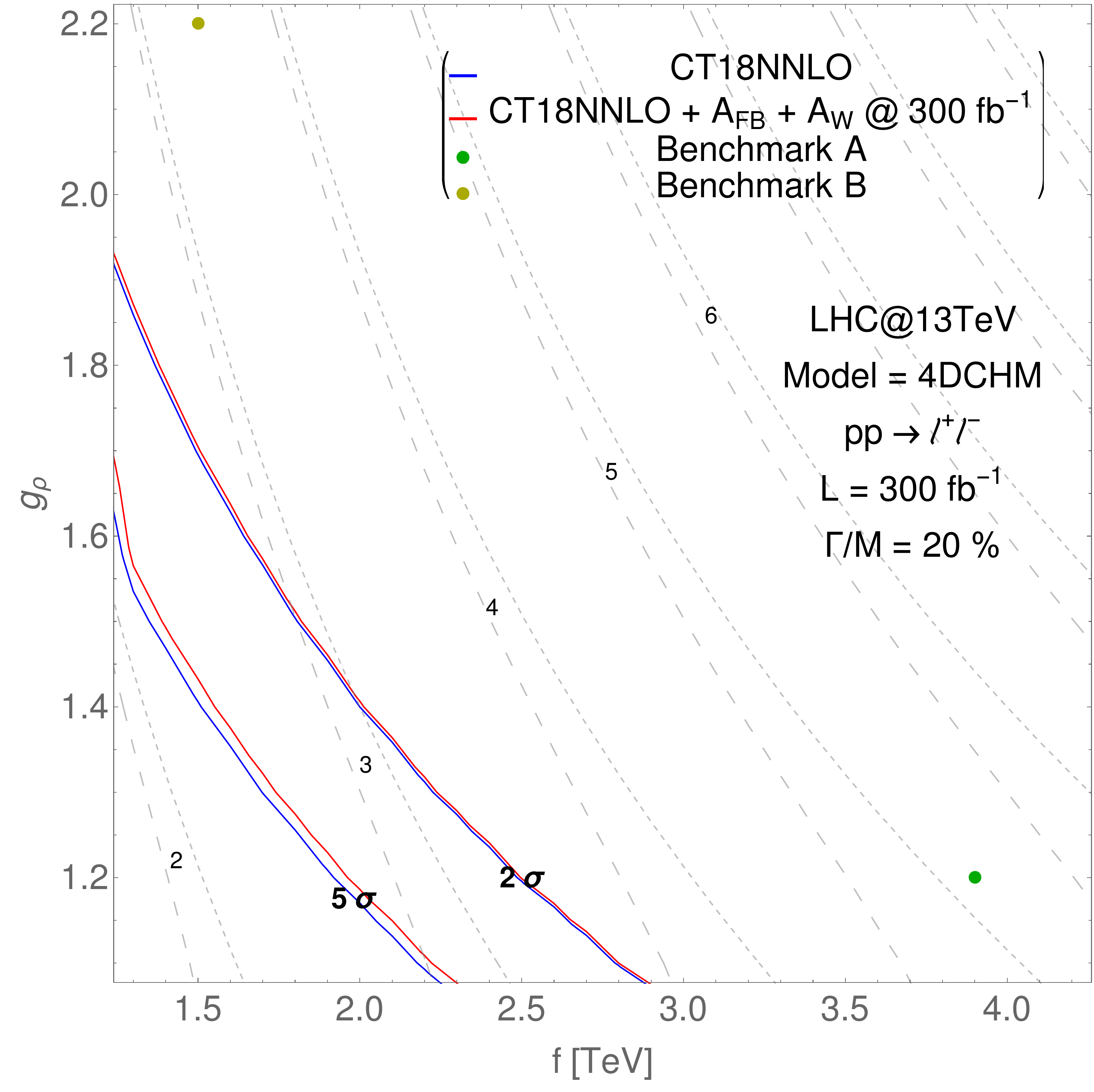}
\includegraphics[width=0.4\textwidth]{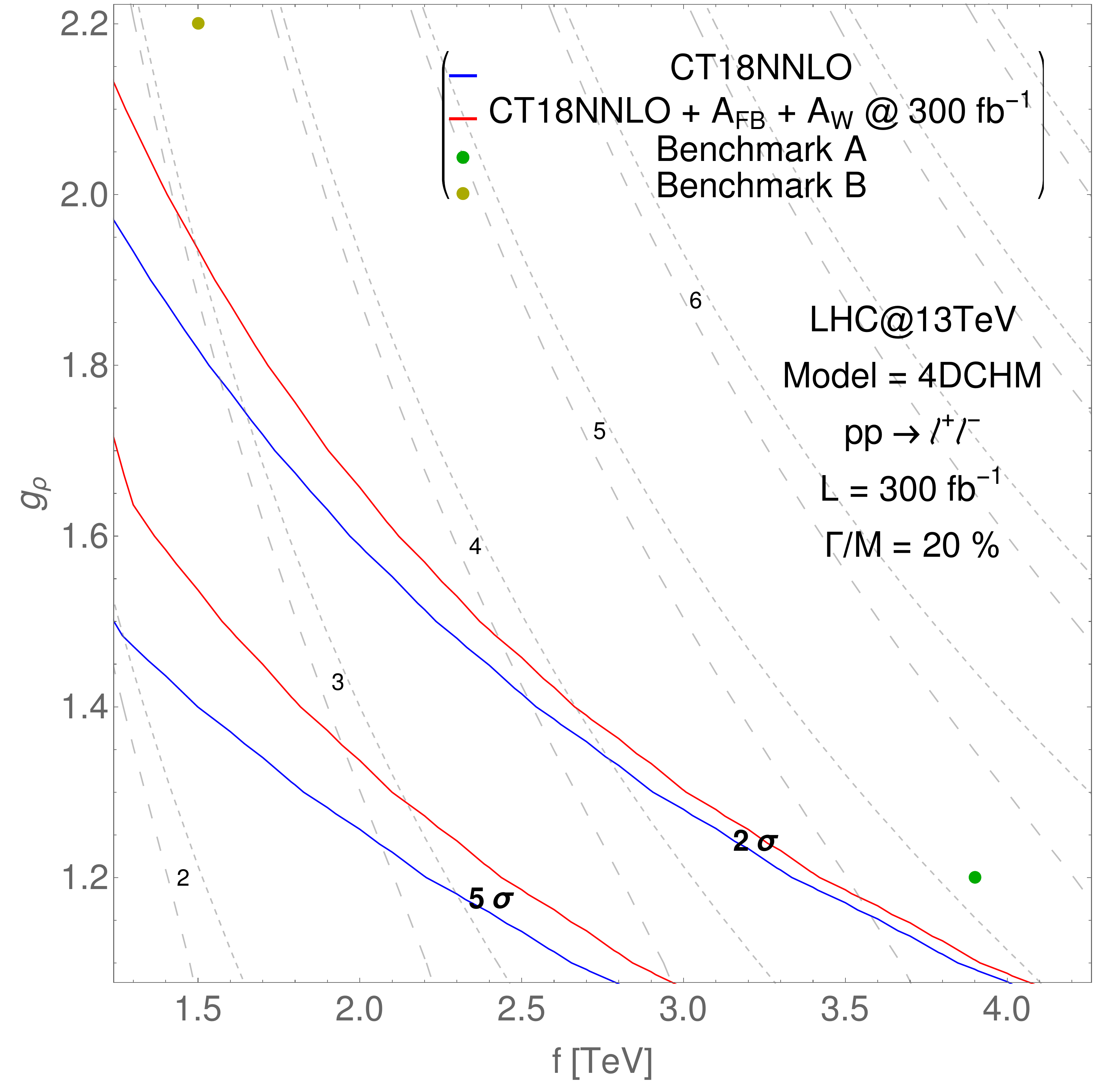}
\end{center}
\caption{Exclusion and discovery limits at 300 fb$^{-1}$ for the peak (left) and for the dip (right) for $Z^{\prime}$ resonances with $\Gamma_{Z^\prime} / M_{Z^\prime} =$ 20\%.
The short (long) dashed contours give the boson mass $M_{Z_2}$ ($M_{Z_3} \simeq M_{W_2}$) in TeV.}
\label{fig:Contour_20_300}
\end{figure}

\subsection{Limits for LHC Run-III}

In Fig.~\ref{fig:Contour_20_300}, we show the limits on the model parameter space for the LHC Run 3 with Centre-of-Mass (CM) energy of 13 TeV and an integrated luminosity of 300 fb$^{-1}$.
We present the results for the peak (left-hand side) and for the dip (right-hand side) in the multiresonant profile. 
For better reading of the results, in the background we also give the contour plots for the masses of the gauge bosons, $M_{Z_2}$ (short-dashed curves) and $M_{Z_3} \simeq M_{W_2}$ (long-dashed curves).
We see that, with the base PDFs, the sensitivity to the dip in comparison with the one from the peak is much higher in the parameter space region with large $f$ and small $g_\rho$, while it is only slightly higher for exclusion in the region with small $f$ and large $g_\rho$, but not for discovery.

This effect is related to the profiles of the invariant mass distributions occurring in the two parameters space regions.
As already shown in Ref.~\cite{Accomando:2016mvz}, for small values of the composite scale $f$ the two neutral resonances $Z_2$ and $Z_3$ are very close in mass, thus their interference effects (in particular with the SM gauge bosons) accumulate in a confined invariant mass region below the peak, and furthermore they get stronger when the coupling $g_\rho$ is large, resulting in a narrow, well pronounced dip.
In the complementary parameter space region (i.e., at large $f$) the BSM neutral resonances are well separated in mass and the contribution of their interference effects with the SM background is spread over a wider invariant mass interval, resulting in a broad, less pronounced dip.
As expected, the amelioration of the PDF error due to the profiled PDFs is critical in the former model's parameter region and marginal in the latter, with the overall result that the dip selection is vastly more constraining than the peak one.
Finally, as visible in this setup, the selected benchmarks, denoted by the red dots in the $(f, g_\rho)$ parameter plane of 
Fig.~\ref{fig:Contour_20_300}, would be far below the experimental sensitivity.

\subsection{Limits for HL-LHC}

\begin{figure}
\begin{center}
\includegraphics[width=0.4\textwidth]{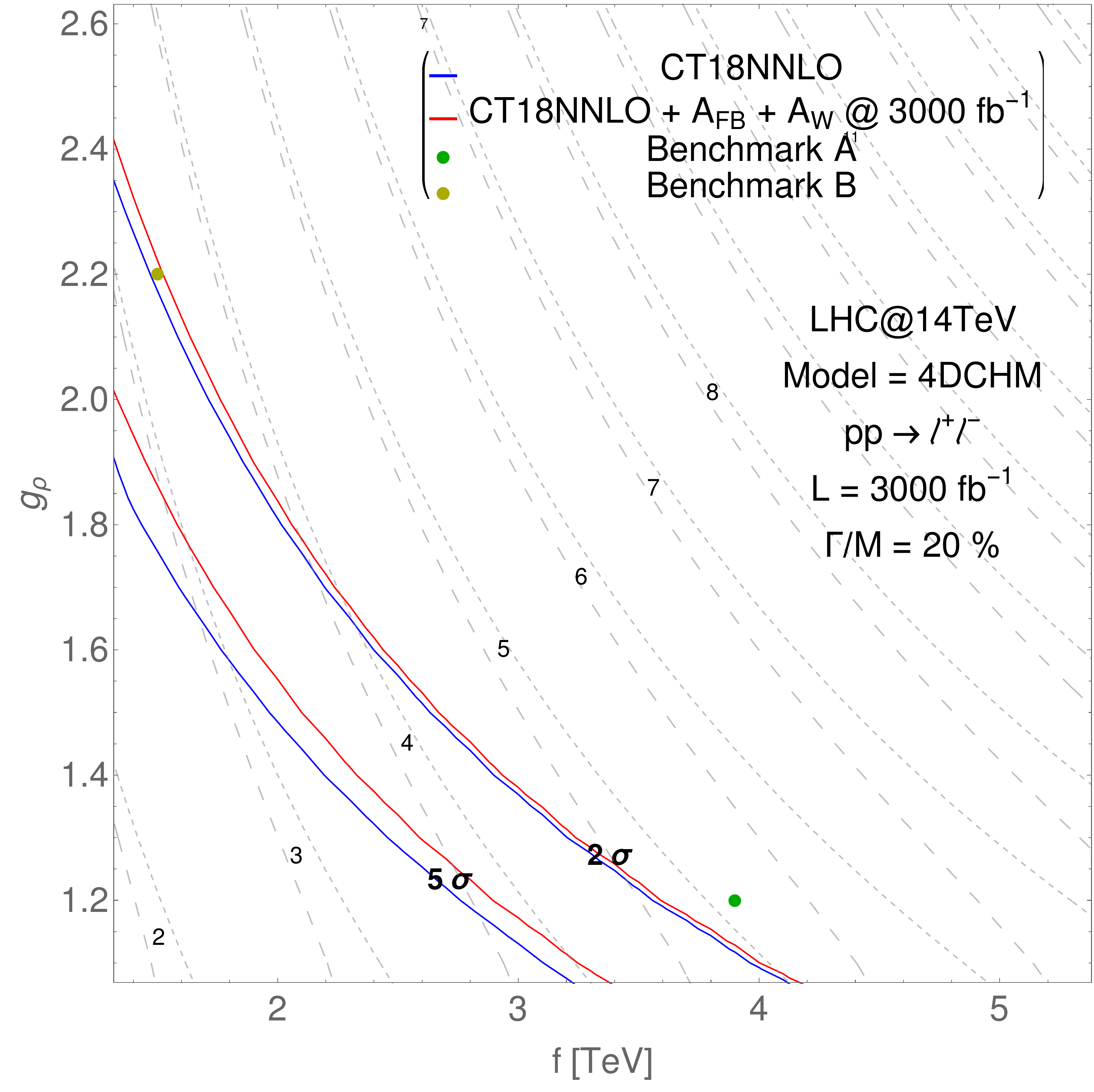}
\includegraphics[width=0.4\textwidth]{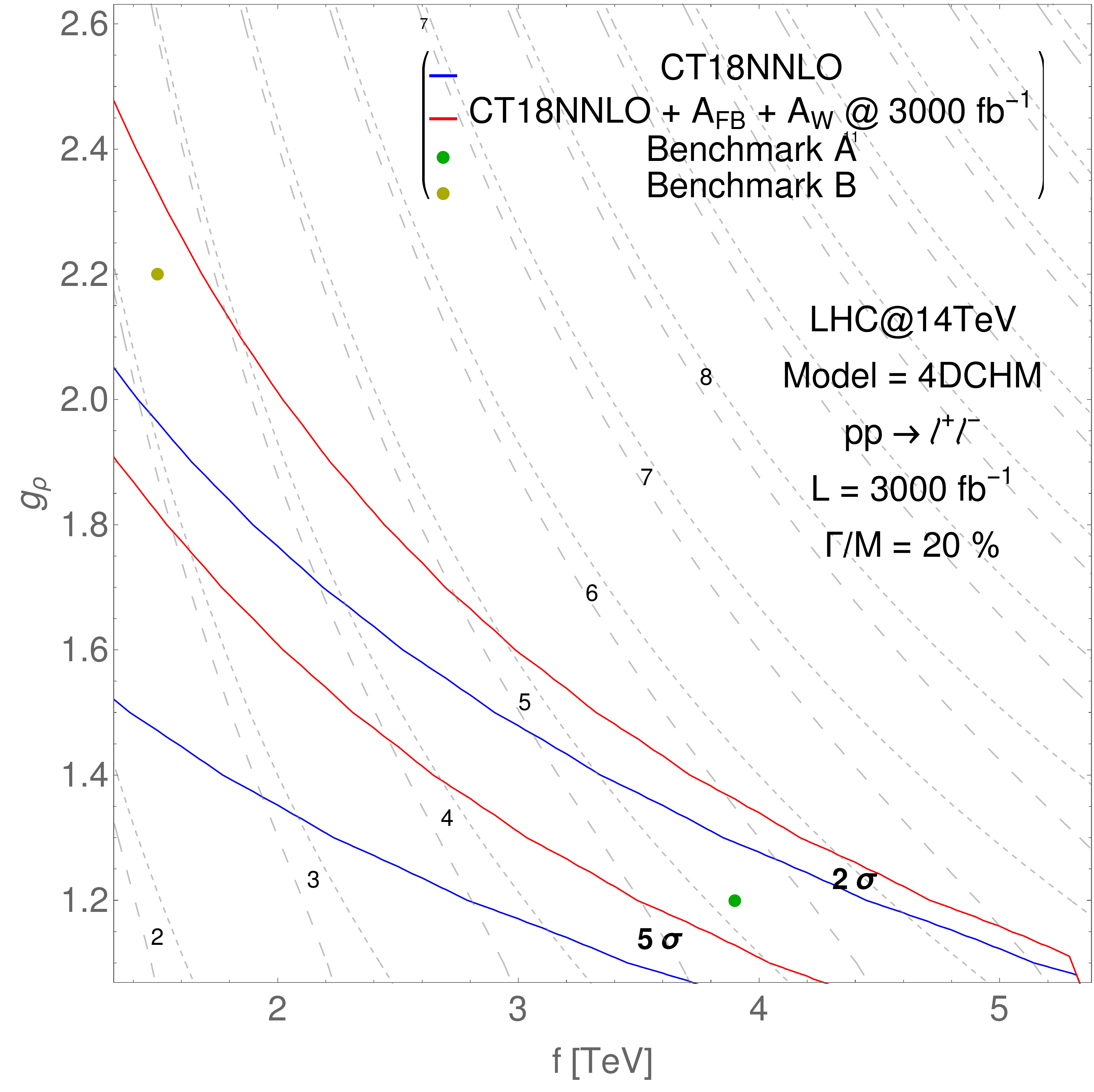}
\end{center}
\caption{Exclusion and discovery limits at 3000 fb$^{-1}$ for the peak (left) and for the dip (right) for $Z^{\prime}$ resonances with $\Gamma / M$ = 20\%.
The short (long) dashed contours give the boson mass $M_{Z_2}$ ($M_{Z_3} \simeq M_{W_2}$) in TeV.}
\label{fig:Contour_20_3000}
\end{figure}

In Fig.~\ref{fig:Contour_20_3000}, we show the limits on the model parameter space for the HL-LHC stage with CM energy of 14 TeV and 
an integrated luminosity of 3000 fb$^{-1}$, with mass contour plots in the background. In this setup, the peak of benchmark A would 
still be below the experimental sensitivity, while the peak of benchmark B, if the improved PDFs are employed, would be right below the 2$\sigma$ exclusion.
When exploiting the depletion of events in the dip below the peak, the sensitivity on the model increases remarkably.
Furthermore, as visible from the right plot of Fig.~\ref{fig:Contour_20_3000}, the improvement on the PDF also has a very large impact particularly in the region of small $f$ and large $g_\rho$, where the sensitivity on the dip can overtake the LHC reach in the peak region once the profiled PDFs are employed. 
Taking into account the reduction of PDF uncertainty, benchmark A would be now at the edge of the 5$\sigma$ discovery, while the sensitivity on benchmark B would almost reach 3$\sigma$.

\begin{figure}
\begin{center}
\includegraphics[width=0.4\textwidth]{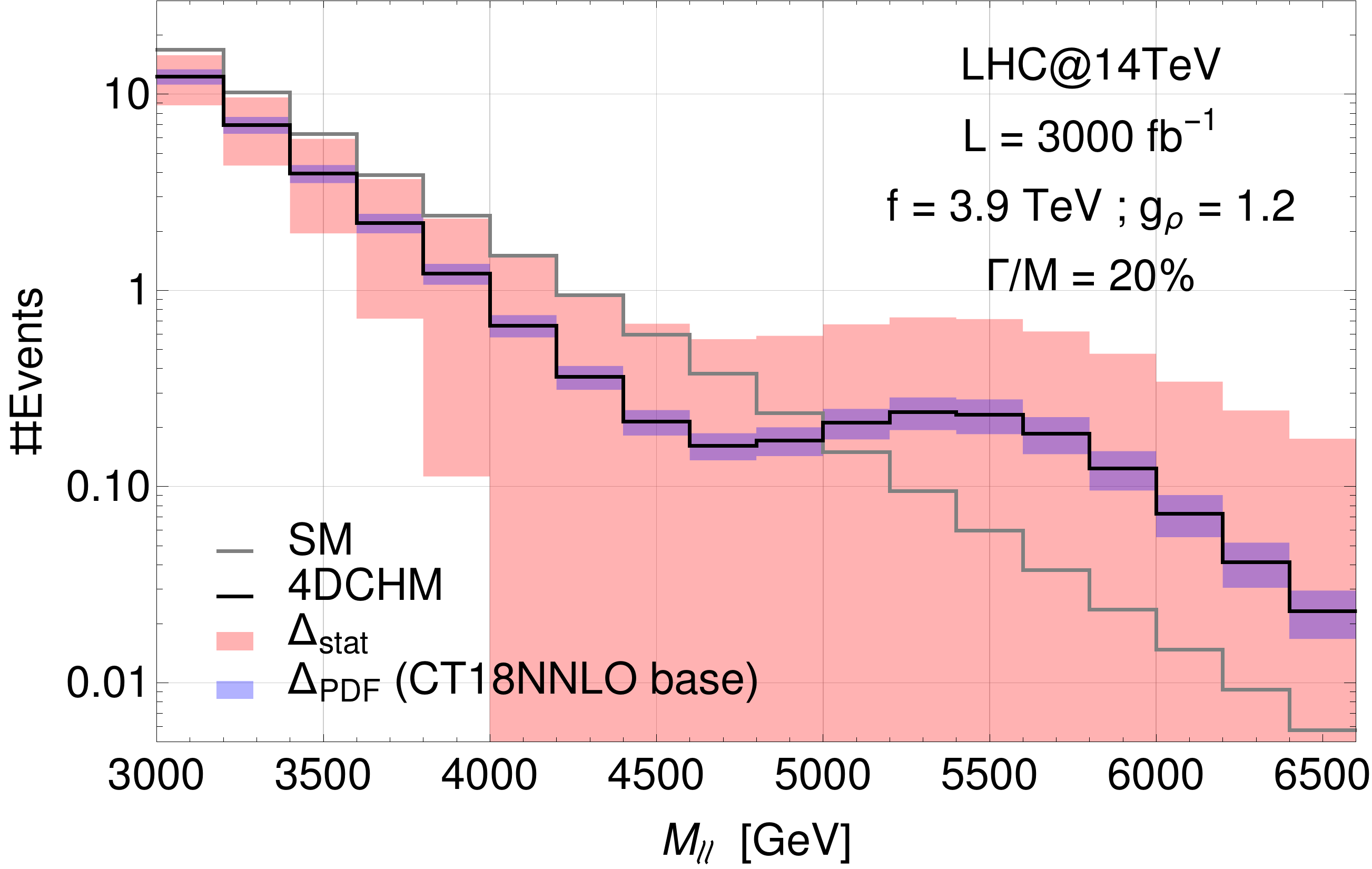}
\includegraphics[width=0.4\textwidth]{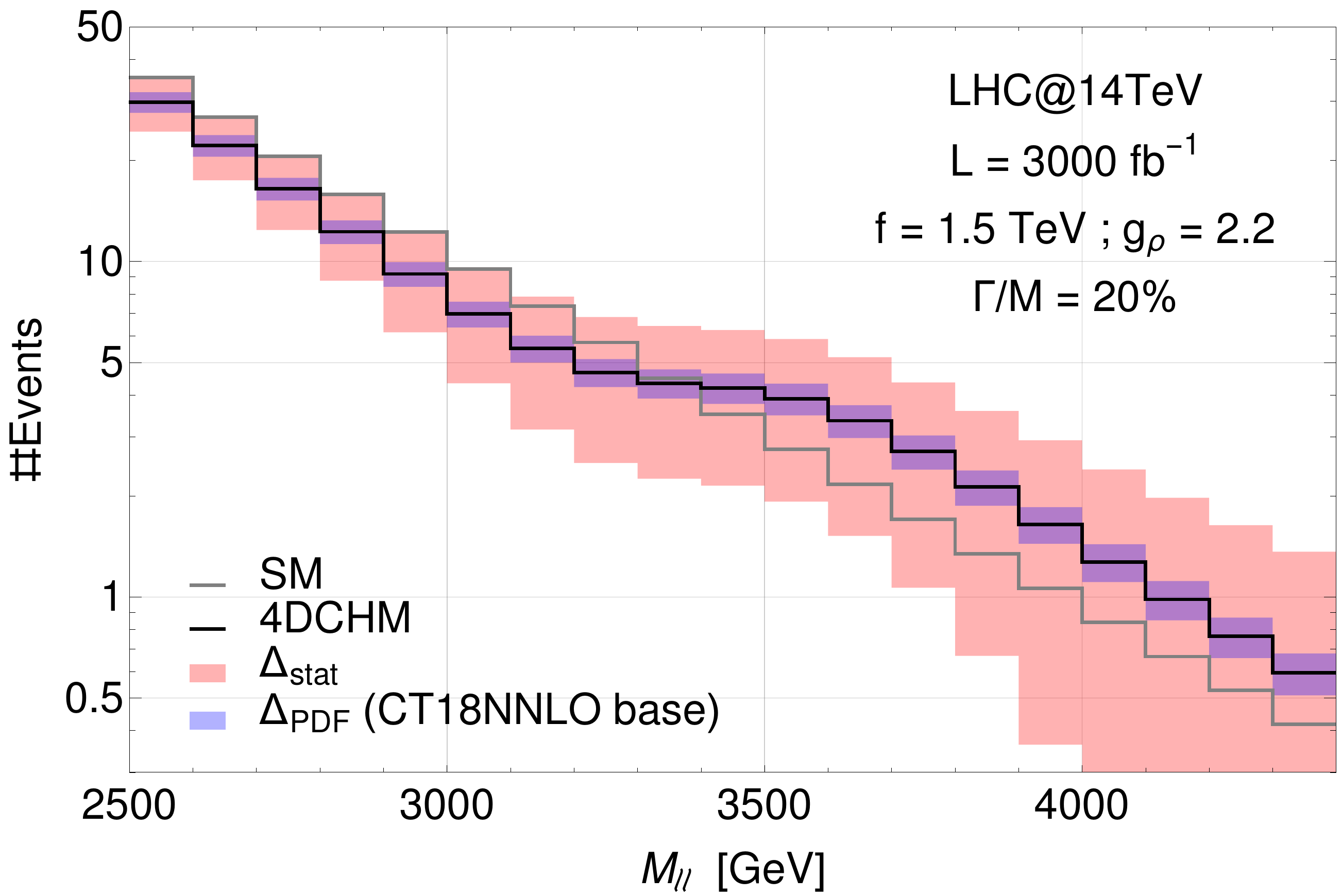}
\end{center}
\caption{Number of events as function of the di-lepton invariant mass for benchmark A (left) and benchmark B (right) with 3000 fb$^{-1}$ of integrated luminosity.}
\label{fig:BenchAB_neutral}
\end{figure}

In such circumstances, it is worthwhile to explore the potential for evidence or discovery at the HL-LHC. We do so in Fig.~\ref{fig:BenchAB_neutral}, where we show the distribution of the number of events as a function of the di-lepton invariant mass at an integrated luminosity of 3000 fb$^{-1}$, for the two benchmarks. The reduction of PDF error turns out to be relevant, even if the statistical error dominates.

This is made clear in 
Tab.~\ref{tab:AB_peak_neutral}, in which we list, in the case of the two benchmarks A and B, the integration intervals in invariant mass for the peak region as well as the resulting cross sections for the SM background and the complete 4DCHM, together with the associated PDF uncertainty using the baseline CT18NNLO PDF set~\cite{Hou:2019efy} and the profiled version~\cite{Fiaschi:2021okg} using the $A_{\rm{FB}}$ and $A_W$ combination with 3000 fb$^{-1}$ of integrated luminosity. We report also the obtained significances for an integrated luminosity of 3000 fb$^{-1}$ employing the two PDF errors. Equivalent results for the dip region are reported in Tab.~\ref{tab:AB_dip_neutral}, with the corresponding choices of integration intervals. While the improvement in significance due to the profiles PDFs is more marginal in the first case, between 3\% and 10\%, it becomes sizeable in the second case, as here significances grow by an amount from 40\% to 90\% (i.e., by a factor of about 10 in comparison) for the benchmarks A and B, respectively.

\begin{table}
\begin{center}

\begin{tabular}{|c|c|c|c|}
\hline
\multicolumn{4}{|c|}{Benchmark A}\\
\hline
inf [TeV] & sup [TeV] & $\sigma_{\rm SM}$ [fb] & $\sigma_{\rm SM+BSM}$ [fb] \\
4.99 & 8.90 & 1.36 $\cdot$ 10$^{-4}$ & 3.87 $\cdot$ 10$^{-4}$ \\
\hline
$\Delta_{\rm PDF}$ base [fb] & $\Delta_{\rm PDF}$ profiled [fb] & $\alpha$ (base) & $\alpha$ (profiled)\\
8.1 $\cdot$ 10$^{-5}$ & 5.6 $\cdot$ 10$^{-5}$ & 1.31 & 1.35 \\
\hline
\end{tabular}
\begin{tabular}{|c|c|c|c|}
\hline
\multicolumn{4}{|c|}{Benchmark B}\\
\hline
inf [TeV] & sup [TeV] & $\sigma_{\rm SM}$ [fb] & $\sigma_{\rm SM+BSM}$ [fb] \\
3.36 & 5.52 & 5.97 $\cdot$ 10$^{-3}$ & 8.34 $\cdot$ 10$^{-3}$ \\
\hline
$\Delta_{\rm PDF}$ base [fb] & $\Delta_{\rm PDF}$ profiled [fb] & $\alpha$ (base) & $\alpha$ (profiled)\\
9.9 $\cdot$ 10$^{-4}$ & 5.8 $\cdot$ 10$^{-4}$ & 1.88 & 2.10 \\
\hline
\end{tabular}
\end{center}
\caption{Integration limits for the peak region, integrated cross section for the SM background and the complete model and its PDF uncertainty with the baseline CT18NNLO PDF set~\cite{Hou:2019efy} and the profiled PDF set using $A_{\rm FB} + A_W$ pseudodata as well as the significances $\alpha$ employing the two PDF errors for an integrated luminosity of 3000 fb$^{-1}$ for the benchmarks A and B.}
\label{tab:AB_peak_neutral}
\end{table}

\begin{table}
\begin{center}

\begin{tabular}{|c|c|c|c|}
\hline
\multicolumn{4}{|c|}{Benchmark A}\\
\hline
inf [TeV] & sup [TeV] & $\sigma_{\rm SM}$ [fb] & $\sigma_{\rm SM+BSM}$ [fb] \\
2.06 & 4.99 & 1.69 $\cdot$ 10$^{-1}$ & 1.42 $\cdot$ 10$^{-1}$ \\
\hline
$\Delta_{\rm PDF}$ base [fb] & $\Delta_{\rm PDF}$ profiled [fb] & $\alpha$ (base) & $\alpha$ (profiled)\\
9.5 $\cdot$ 10$^{-3}$ & 4.6 $\cdot$ 10$^{-3}$ & 3.34 & 4.82 \\
\hline
\end{tabular}
\begin{tabular}{|c|c|c|c|}
\hline
\multicolumn{4}{|c|}{Benchmark B}\\
\hline
inf [TeV] & sup [TeV] & $\sigma_{\rm SM}$ [fb] & $\sigma_{\rm SM+BSM}$ [fb] \\
1.36 & 3.36 & 1.53 & 1.45 \\
\hline
$\Delta_{\rm PDF}$ base [fb] & $\Delta_{\rm PDF}$ profiled [fb] & $\alpha$ (base) & $\alpha$ (profiled)\\
6.8 $\cdot$ 10$^{-2}$ & 3.1 $\cdot$ 10$^{-2}$ & 1.53 & 2.91 \\
\hline
\end{tabular}
\end{center}
\caption{Integration limits for the dip region, integrated cross section for the SM background and the complete model and its PDF uncertainty with the baseline CT18NNLO PDF set~\cite{Hou:2019efy} and the profiled PDF set using $A_{\rm FB} + A_W$ pseudodata as well as the significances $\alpha$ employing the two PDF errors for an integrated luminosity of 3000 fb$^{-1}$ for the benchmarks A and B.}
\label{tab:AB_dip_neutral}
\end{table}

\section{Limits on the model parameter space from Charged DY}
\label{sec:CC}

We will see in this section that the CC channel in general provides stronger bounds on the model parameter space.
The 4DCHM spectrum in the charged sector accounts for two active $W^\prime$ bosons, $W_2$ and $W_3$~\cite{Barducci:2012kk}.
The two resonances are generally well separated in mass, as in the parameter space region of interest which we are exploring we have $M_{W_3} - M_{W_2} \simeq \mathcal{O}$(1 TeV).
The $W_3$ boson is thus heavy and weakly coupled to the SM matter, so that its phenomenology is not relevant in terms of BSM searches at colliders.

Similarly to the analysis in the NC channel, due to the typical Jacobian peak shape of the transverse mass distribution of the decay products, we will consider again a ``counting strategy" analysis, choosing optimal integration limits to maximize the sensitivity to the BSM realization adopted here. Furthermore, 
as previously, we set $\Gamma_{W'}/M_{W'}=20\%$ and the overall significance is again evaluated combining the sensitivities of the final states with electrons or muons, including their acceptances and efficiencies~\cite{CMS:2020cph}. Finally, the 
statistical and PDF uncertainties are summed in quadrature and no other source of systematic error is assumed here either.

\subsection{Limits for LHC Run-III}

\begin{figure}
\begin{center}
\includegraphics[width=0.4\textwidth]{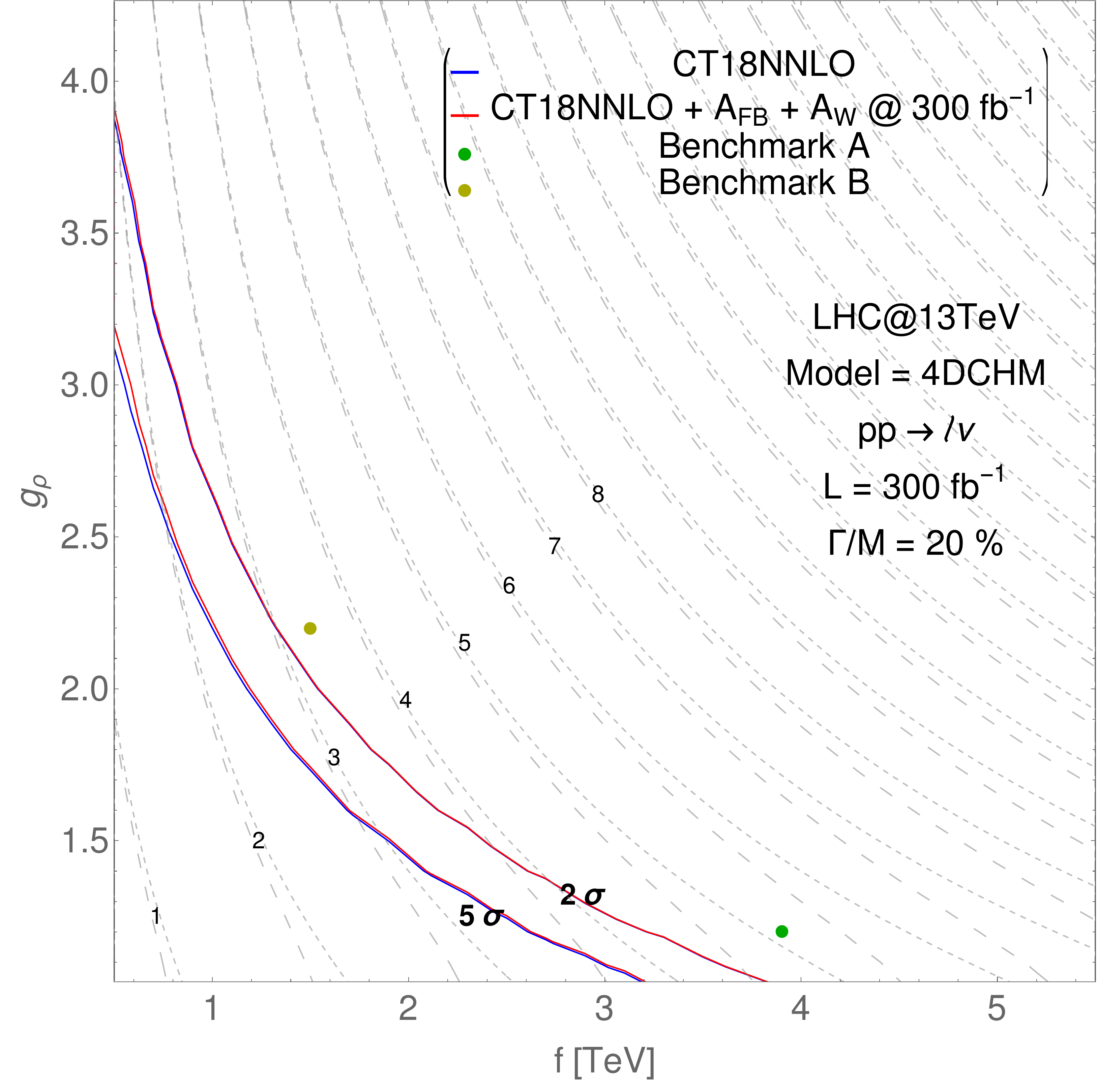}
\includegraphics[width=0.4\textwidth]{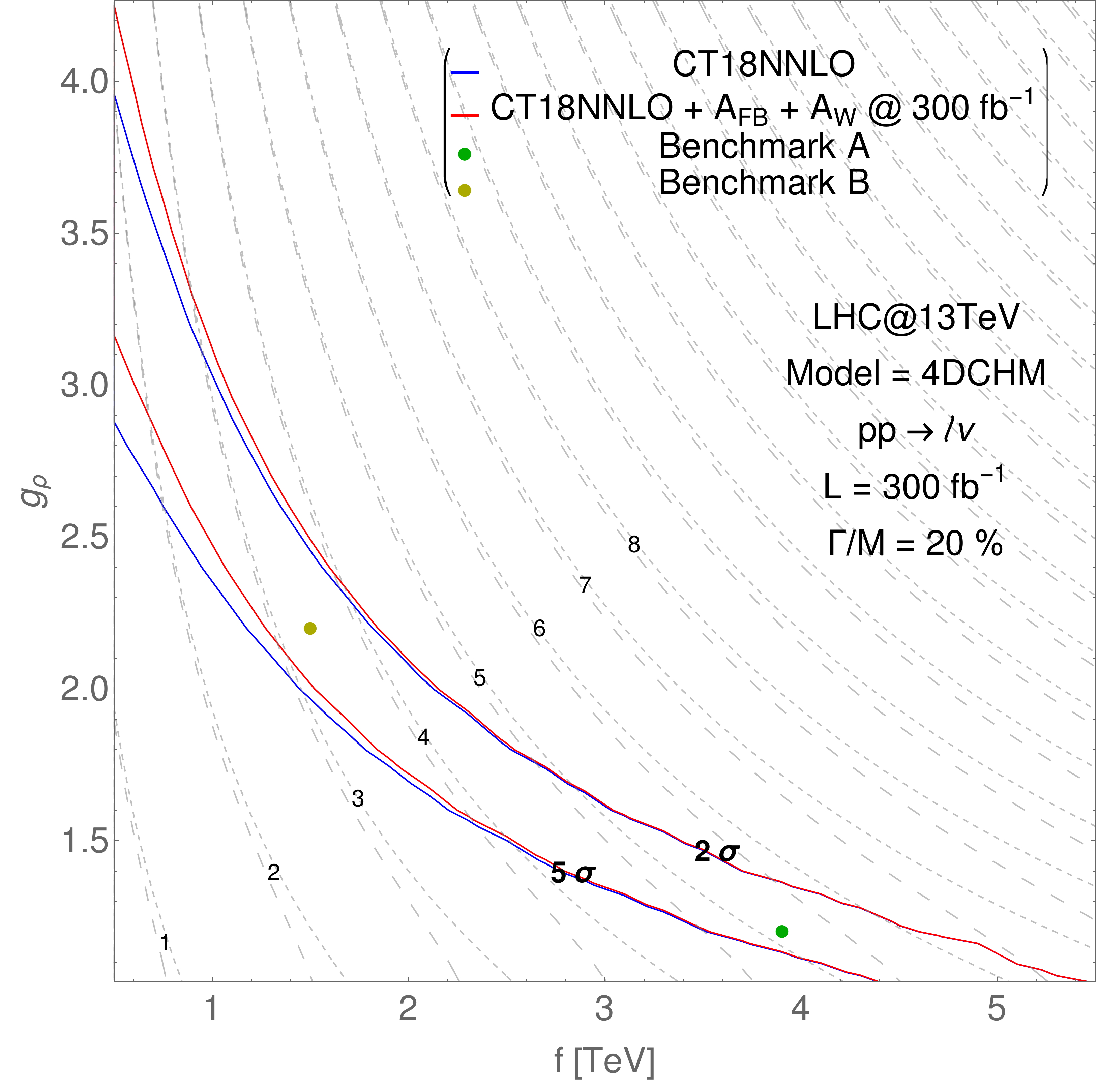}
\end{center}
\caption{Exclusion and discovery limits at 300 fb$^{-1}$ for the peak (left) and for the dip (right) for $W^{\prime}$ resonances with $\Gamma / M$ = 20\%.
The short (long) dashed contours give the boson mass $M_{Z_2}$ ($M_{Z_3} \simeq M_{W_2}$) in TeV.}
\label{fig:Contour_Wpr_20_300}
\end{figure}

In Fig.~\ref{fig:Contour_Wpr_20_300}, we show the limits on the model parameter space for the LHC Run 3 with CM energy of 13 TeV and 
an integrated luminosity of 300 fb$^{-1}$.
The contour plots for the masses of the extra (active) neutral and charged gauge bosons are shown in the background. 
As intimated earlier, and now visible from this plot, the CC channel provides more stringent limits on the 4DCHM, as the production cross section of charged gauge bosons is higher than that of the neutral ones.
On the other hand, as we are dealing with a single BSM resonance (the $W_3$ boson being heavier and less coupled), the interference effects that generate the dip are 
now weaker.
The sensitivity to the dip significantly exceeds that of the peak only in the large-$f$ region, with small improvement due to the reduction of PDF uncertainty.
Still, in the complementary region with small $f$ values, the LHC sensitivity to the model from the peak signal can be slightly extended by the analysis of the dip provided that the profiled PDFs with reduced uncertainty are employed. 
In this setup, the peaks of the two benchmarks still remain below 2$\sigma$.
In contrast, exploiting the depletion of events below the Jacobian peaks, both benchmarks would reach a 3$\sigma$ significance.

\subsection{Limits for HL-LHC}

\begin{figure}
\begin{center}
\includegraphics[width=0.4\textwidth]{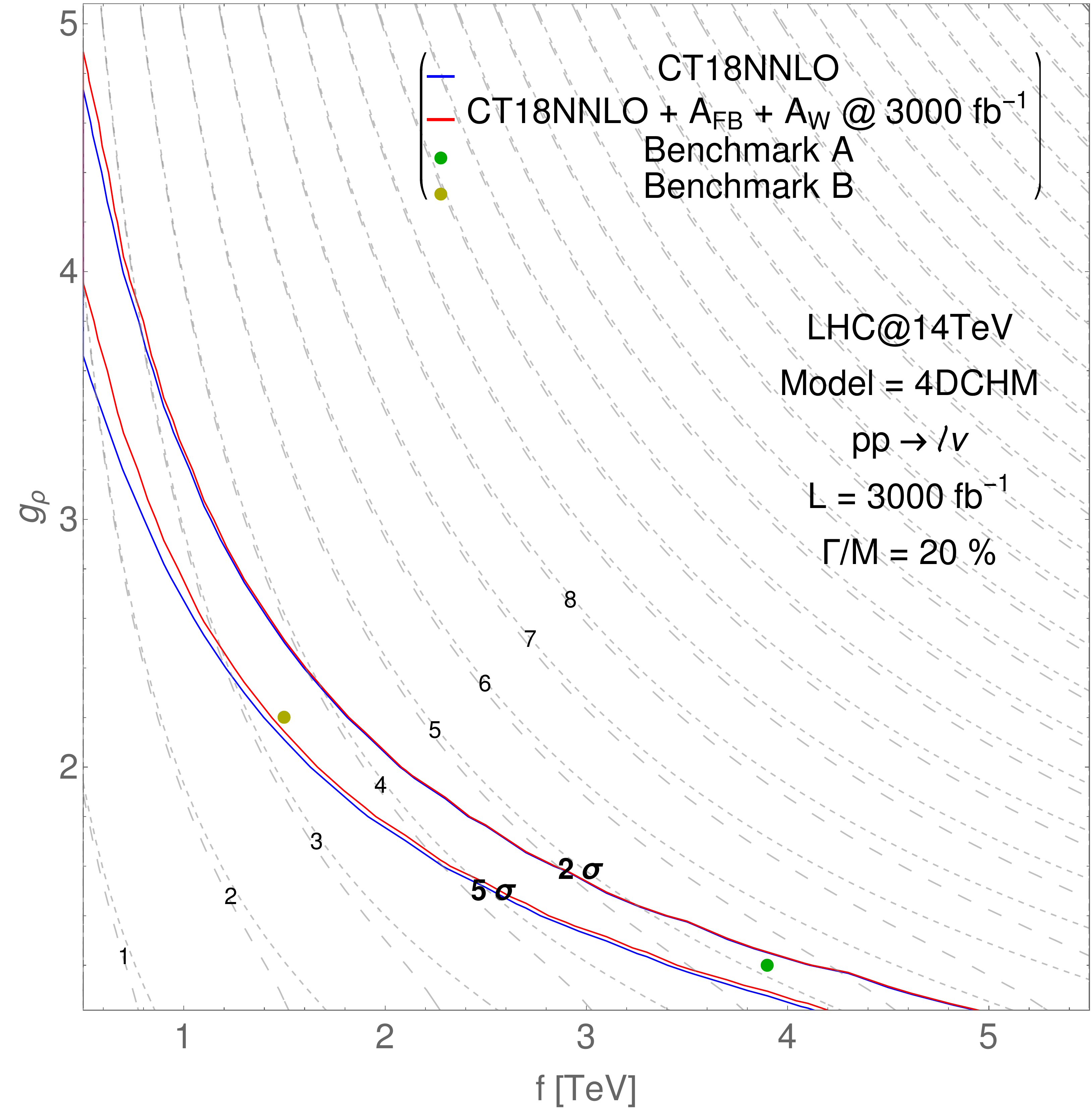}
\includegraphics[width=0.4\textwidth]{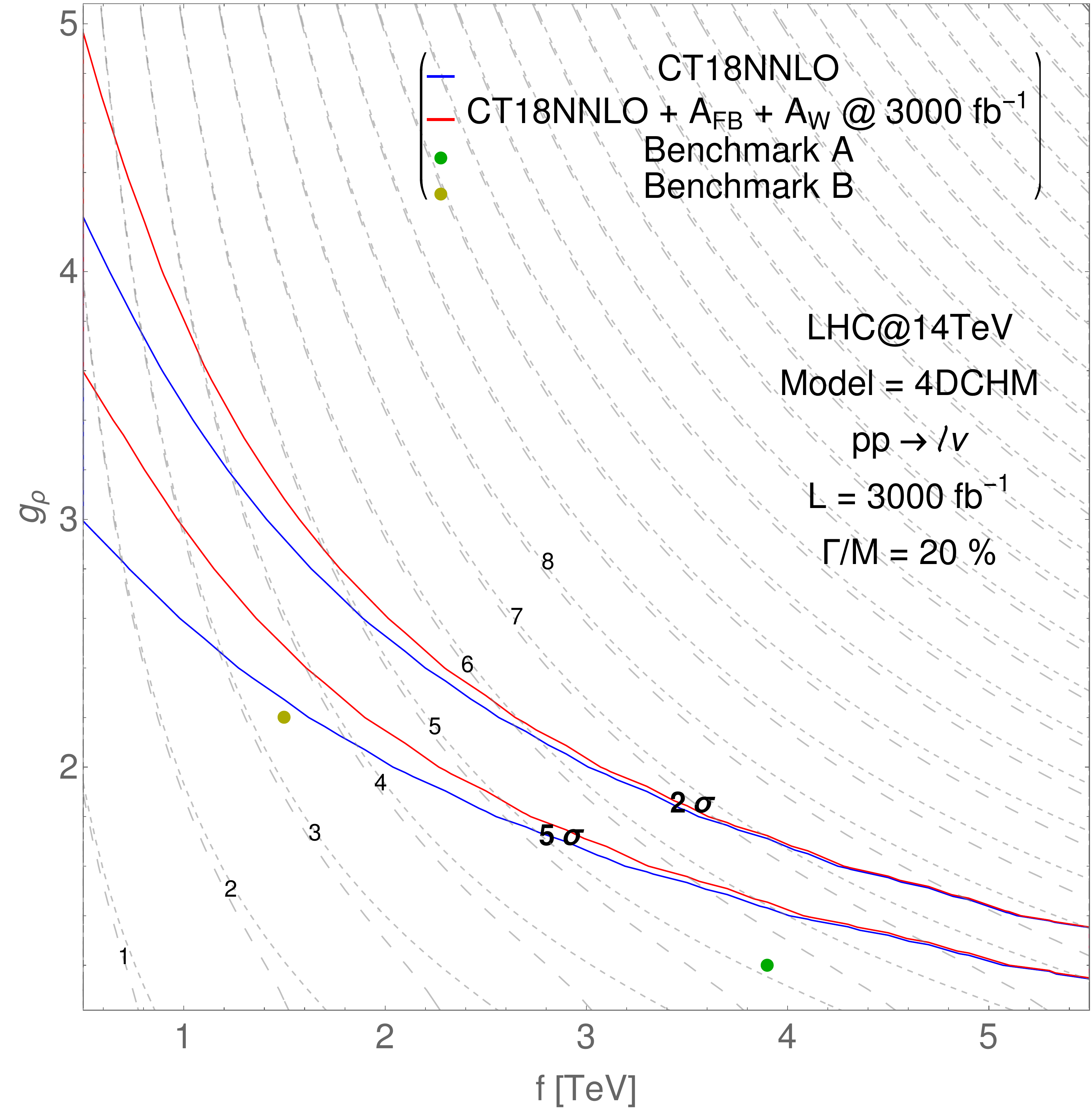}
\end{center}
\caption{Exclusion and discovery limits at 3000 fb$^{-1}$ for the peak (left) and for the dip (right) for $W^{\prime}$ resonances with $\Gamma / M$ = 20\%.
The short (long) dashed contours give the boson mass $M_{Z_2}$ ($M_{Z_3} \simeq M_{W_2}$) in TeV.}
\label{fig:Contour_Wpr_20_3000}
\end{figure}

In Fig.~\ref{fig:Contour_Wpr_20_3000}, we show the limits on the model parameter space for the HL-LHC stage with CM energy of 14 TeV and an integrated luminosity of 3000 fb$^{-1}$.
Similarly to the previous results, we note a significant improvement in the sensitivity to the model from the analysis of the dip in comparison with the analysis of the peak for the region with large values of the compositeness scale $f$.
In the complementary region with small values of $f$, the signal of the peak generally has a larger significance, particularly for discovery purposes. 
The analysis of the dip, however, can be competitive in drawing exclusions, especially when the profiled PDFs are employed.
In this setup, the two benchmarks A and B are both within HL-LHC reach, particularly so in the presence of profiled PDFs.

\begin{figure}
\begin{center}
\includegraphics[width=0.4\textwidth]{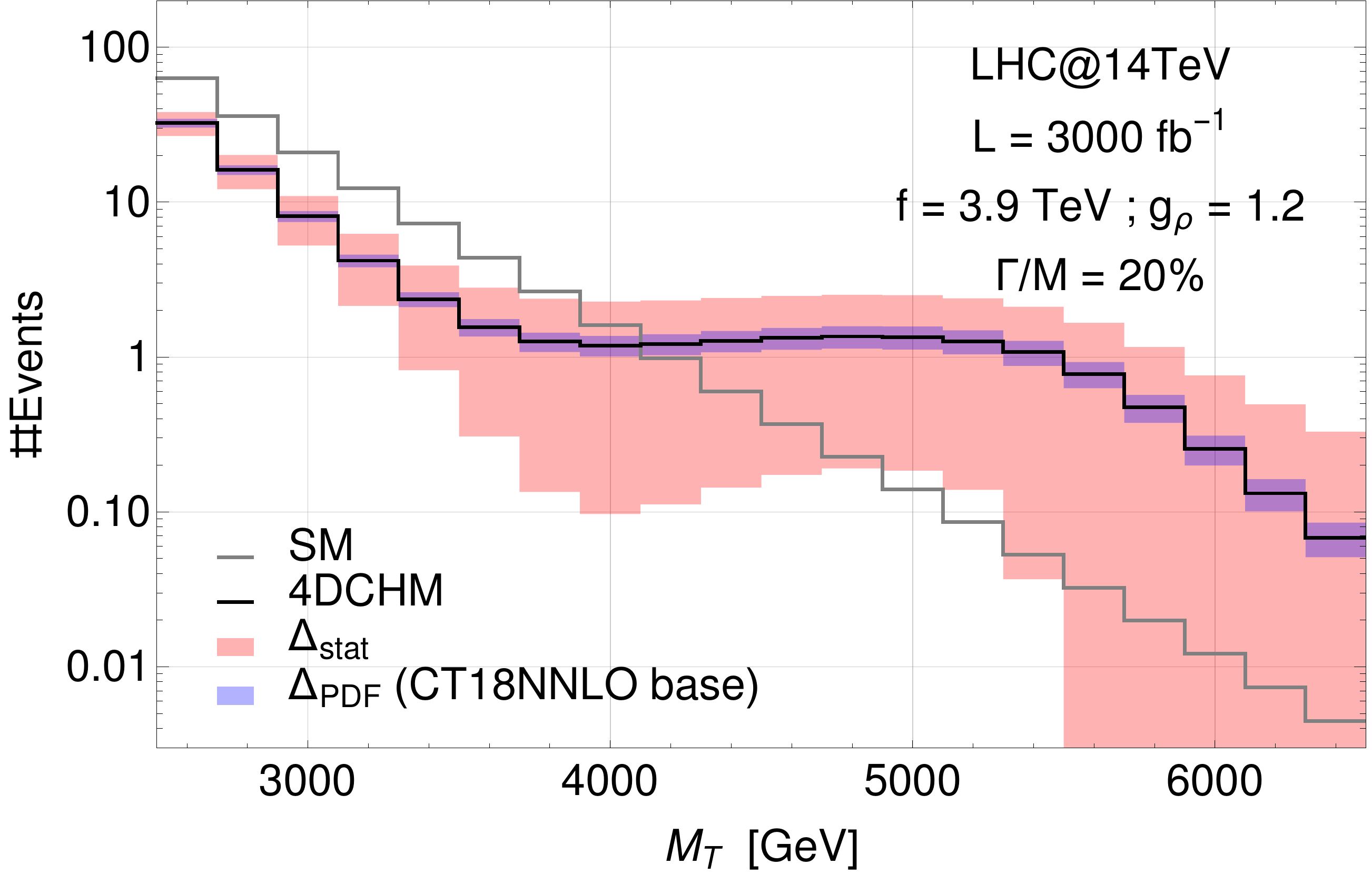}
\includegraphics[width=0.4\textwidth]{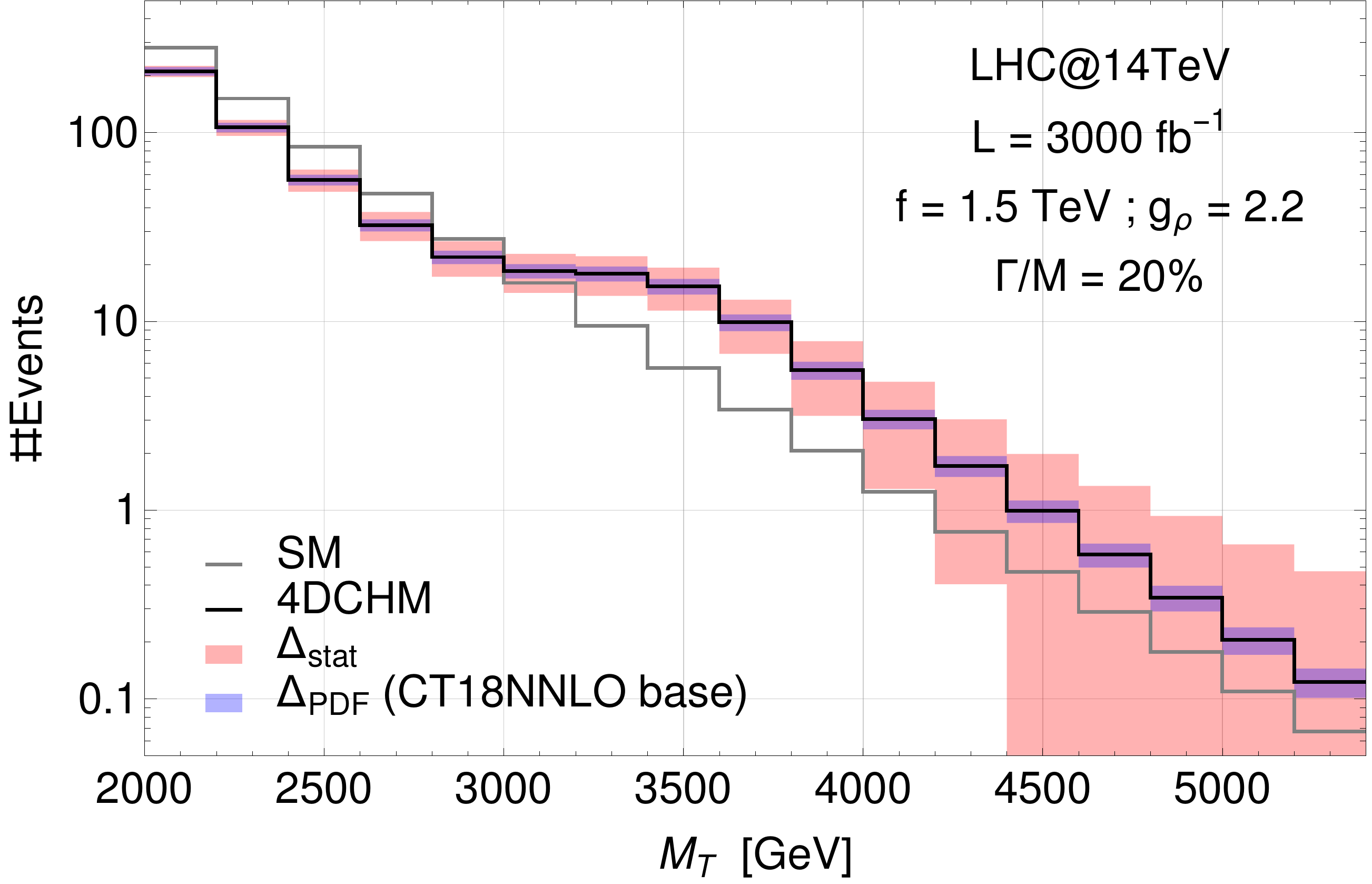}
\end{center}
\caption{Number of events as function of the system transverse mass for benchmark A (left) and benchmark B (right) with 3000 fb$^{-1}$ of integrated luminosity.}
\label{fig:BenchAB_charged}
\end{figure}

Hence, we study again the HL-LHC potential to test the possibility of the existence of a $W'$ state of the 4DCHM. Specifically, in Fig.~\ref{fig:BenchAB_charged}, we show the distribution of the number of events as a 
function of the transverse mass of the lepton-neutrino decay products of the $W'$ at an integrated luminosity of 3000 fb$^{-1}$, for the two benchmarks A and B.

\begin{table}
\begin{center}

\begin{tabular}{|c|c|c|c|}
\hline
\multicolumn{4}{|c|}{Benchmark A}\\
\hline
inf [TeV] & sup [TeV] & $\sigma_{\rm SM}$ [fb] & $\sigma_{\rm SM+BSM}$ [fb] \\
4.11 & 8.90 & 8.13 $\cdot$ 10$^{-4}$ & 3.51 $\cdot$ 10$^{-3}$ \\
\hline
$\Delta_{\rm PDF}$ base [fb] & $\Delta_{\rm PDF}$ profiled [fb] & $\alpha$ (base) & $\alpha$ (profiled)\\
6.1 $\cdot$ 10$^{-4}$ & 5.3 $\cdot$ 10$^{-4}$ & 2.69 & 2.75 \\
\hline
\end{tabular}

\begin{tabular}{|c|c|c|c|}
\hline
\multicolumn{4}{|c|}{Benchmark B}\\
\hline
inf [TeV] & sup [TeV] & $\sigma_{\rm SM}$ [fb] & $\sigma_{\rm SM+BSM}$ [fb] \\
3.03 & 5.52 & 1.22 $\cdot$ 10$^{-2}$ & 2.36 $\cdot$ 10$^{-2}$ \\
\hline
$\Delta_{\rm PDF}$ base [fb] & $\Delta_{\rm PDF}$ profiled [fb] & $\alpha$ (base) & $\alpha$ (profiled)\\
2.3 $\cdot$ 10$^{-3}$ & 1.9 $\cdot$ 10$^{-3}$ & 4.00 & 4.24 \\
\hline
\end{tabular}
\end{center}
\caption{Integration limits for the peak region, integrated cross section for the SM background and the complete model and its PDF uncertainty with the baseline CT18NNLO PDF set~\cite{Hou:2019efy} and the profiled PDF set using $A_{\rm FB} + A_W$ pseudodata as well as the significances $\alpha$ employing the two PDF errors for an integrated luminosity of 3000 fb$^{-1}$ for the benchmarks A and B.}
\label{tab:AB_peak_charged}
\end{table}

\begin{table}
\begin{center}

\begin{tabular}{|c|c|c|c|}
\hline
\multicolumn{4}{|c|}{Benchmark A}\\
\hline
inf [TeV] & sup [TeV] & $\sigma_{\rm SM}$ [fb] & $\sigma_{\rm SM+BSM}$ [fb] \\
2.22 & 4.11 & 1.07 $\cdot$ 10$^{-1}$ & 5.71 $\cdot$ 10$^{-2}$ \\
\hline
$\Delta_{\rm PDF}$ base [fb] & $\Delta_{\rm PDF}$ profiled [fb] & $\alpha$ (base) & $\alpha$ (profiled)\\
3.7 $\cdot$ 10$^{-3}$ & 2.7 $\cdot$ 10$^{-3}$ & 11.16 & 12.21 \\
\hline
\end{tabular}

\begin{tabular}{|c|c|c|c|}
\hline
\multicolumn{4}{|c|}{Benchmark B}\\
\hline
inf [TeV] & sup [TeV] & $\sigma_{\rm SM}$ [fb] & $\sigma_{\rm SM+BSM}$ [fb] \\
1.38 & 3.03 & 1.60 & 1.36 \\
\hline
$\Delta_{\rm PDF}$ base [fb] & $\Delta_{\rm PDF}$ profiled [fb] & $\alpha$ (base) & $\alpha$ (profiled)\\
5.7 $\cdot$ 10$^{-2}$ & 3.6 $\cdot$ 10$^{-2}$ & 5.51 & 7.89 \\
\hline
\end{tabular}
\end{center}
\caption{Integration limits for the dip region, integrated cross section for the SM background and the complete model and its PDF uncertainty with the baseline CT18NNLO PDF set~\cite{Hou:2019efy} and the profiled PDF set using $A_{\rm FB} + A_W$ pseudodata as well as the significances $\alpha$ employing the two PDF errors for an integrated luminosity of 3000 fb$^{-1}$ for the benchmarks A and B.}
\label{tab:AB_dip_charged}
\end{table}

In Tab.~\ref{tab:AB_peak_charged} we list, in the case of the two benchmarks A and B, the integration intervals in transverse mass for the peak region and the resulting cross sections for the SM background and the complete 4DCHM, together with the associated PDF uncertainty using the baseline CT18NNLO PDF set~\cite{Hou:2019efy} and the profiled PDF set~\cite{Fiaschi:2021okg} using the $A_{\rm{FB}}$ and $A_W$ combination with 3000 fb$^{-1}$ of integrated luminosity. 
We report also the obtained significances employing the two PDF errors.
Equivalent results for the dip region are reported in Tab.~\ref{tab:AB_dip_charged}, with the corresponding choices of integration intervals. In particular, we see that, 
in the HL-LHC setup, the sensitivity to the model is greatly increased, as the peak of benchmarks A and B reach about 3$\sigma$ and 4$\sigma$ deviations. Indeed, the sensitivity to the model from the depletion of events driven by interference effects is in this context predominant, as both benchmarks give deviations $>$ 5$\sigma$ in the low mass tail region. As visible from the right plot of Fig.~\ref{fig:Contour_Wpr_20_3000}, the improvement in the PDFs would benefit the experimental analysis especially in the parameter space region with small $f$ and large $g_\rho$.

\subsection{Comment on combining neutral and charged DY}

Having analyzed searches in the DY neutral and charged sectors, we conclude this discussion with a comment on combining these searches.\footnote{Phenomenological studies based on combined NC and CC searches have been performed in Ref.~\cite{Araz:2021dga} in the context of the Sequential Standard Model and Left-Right Symmetric Models.}
In all previous figures showing the significances over the $(f,g_\rho)$ plane, we have plotted in the background the masses of the additional (active) neutral and charged gauge bosons, i.e., $M_{Z_2}$ and $M_{Z_3}\simeq M_{W_2}$. In the 4DCHM, such masses (and corresponding couplings in our two benchmark points) are correlated.
By comparing Fig.~\ref{fig:Contour_20_300} to Fig.~\ref{fig:Contour_Wpr_20_300} and Fig.~\ref{fig:Contour_20_3000} to Fig.~\ref{fig:Contour_Wpr_20_3000}, we observe that results obtained in $W^\prime$ searches via the CC channel can be used to constrain $Z^\prime$ properties as well, in a manner more stringent than that afforded by direct searches for it in the NC channel. In particular, comparing the left-hand side to the right-hand side in these two pairs of plots, we stress that this is a particularly efficient approach in the case of profiled PDFs, since up to a 1 TeV more sensitivity can indirectly be gained to $Z^\prime$ masses in $W^\prime$ direct searches with respect to the scope of $Z^\prime$ direct searches themselves.
We further remark that while these conclusions have general validity, the precise size of the quantitative improvements from the combination of NC and CC analyses, as well as from the use of profiled PDFs, depends on the specific realisation of the BSM resonances and in particular on their widths.

\section{Conclusions} 
\label{sec:conc} 

We have herein analyzed LHC searches for broad $W^\prime$ and $ Z^\prime$ vector resonances using as theoretical scenario the gauge sector of the 4DCHM, a calculable minimal representation of a CHM. This is a scenario in which such states emerge naturally, owing to the $W^\prime$ and $Z^\prime$ being able to decay into a variety of final states involving new fermionic states emerging from the strong dynamics of the underlying theory.
We have shown that improving the non-perturbative QCD systematic uncertainties associated with the initial state PDFs, according to an approach that we have recently suggested to profile $A_{\rm FB}$ and $A_W$ asymmetry observables using the \texttt{xFitter} framework, enhances significantly the potential of forthcoming LHC experiments both in discovering and in setting exclusion bounds, via CC and NC DY processes, on broad $W^\prime$ and $ Z^\prime$. In fact, we regard this as a generic result applicable to any theoretical model embedding wide $W^\prime$ and/or $Z^\prime$ states. 

For cases in which both two new states appear and are correlated by theory, we emphasize 
that the large reduction of systematic error associated to the modelling of PDF documented here can profitably be applied to combined $W^\prime$ and $Z^\prime$ searches. We have in fact shown that in the 4DCHM a direct $W^\prime$ exclusion (or indeed discovery) achieved in the CC channel can be used indirectly to probe the existence of a $Z^\prime$ better than it can be done with direct searches in the NC channel.

Furthermore, in a well-defined theoretical framework like the 4DCHM, we have also elaborated on the possibility of exploiting model-dependent effects, due to the interference of the heavy gauge bosons with the SM, in order to extend the experimental sensitivity of dedicated BSM searches. In this context, we find that the analysis of the dip often provides more stringent limits than the signal coming from the peak, with the reduction of systematic PDF errors playing a crucial part in this conclusion.
In post-discovery diagnostic stages, the inclusion of such model-dependent analyses, with systematic uncertainties under control, would indeed provide an important tool to disentangle the specific BSM construction underlying the new physics signals.

\section*{Acknowledgements}
SM is supported in part through the NExT Institute and acknowledges funding from the STFC Consolidated Grant ST/L000296/1. The work of JF is supported by STFC under the Consolidated Grant ST/T000988/1. JF and SM acknowledge the use of the IRIDIS High Performance Computing facility, and associated support services, at the University of Southampton, in the completion of this work.


\bibliographystyle{JHEP}
\bibliography{references}


\end{document}